\documentclass[twocolumn,letter]{jpsj3}
\usepackage{graphicx}
\usepackage{dcolumn}
\usepackage{bm}
\usepackage{ulem}
\usepackage{braket}
\usepackage{amsmath}
\usepackage{color}

\title{Magnetic Toroidal Moment under Partial Magnetic Order\\ in Hexagonal Zigzag-Chain Compound Ce$_3$TiBi$_5$}

\author{Satoru Hayami$^1$ and Hiroaki Kusunose$^2$}
\inst{
$^1$Graduate School of Science, Hokkaido University, Sapporo 060-0810, Japan \\
$^2$Department of Physics, Meiji University, Kawasaki 214-8571, Japan \\
}

\abst{
We theoretically investigate an antiferromagnetic structure in a hexagonal zigzag-chain compound Ce$_3$TiBi$_5$, which shows a linear magnetoelectric effect in metals and an unfamiliar temperature dependence of magnetic susceptibility. 
We find a partial magnetic order as a coexistence of a nonmagnetic zigzag chain and staggered antiferromagnetic zigzag chains. 
The latter accompany in-plane magnetic toroidal dipole moments which are responsible for the observed behaviors. 
We show that magnetic frustration arising from competing exchange interactions plays an important role to realize such a partial magnetic order. 
Furthermore, we present the important spin-orbit coupling parameters to induce the physical phenomena driven by the magnetic toroidal dipole moment, such as the linear magnetoelectric effect and nonlinear transport. 
}

\begin{document}
\maketitle

In recent years, the cross-correlated phenomena and quantum transports associated with the magnetic toroidal dipole (MTD) moment have attracted much attention in condensed matter physics~\cite{dubovik1975multipole,dubovik1990toroid,gorbatsevich1994toroidal, Spaldin_0953-8984-20-43-434203,cheong2018broken}. 
Although the MTD moment was mainly studied as a microscopic origin of the linear magnetoelectric (ME) effect~\cite{popov1999magnetic,schmid2001ferrotoroidics,EdererPhysRevB.76.214404} and nonreciprocal magneto-optics~\cite{Sawada_PhysRevLett.95.237402,Miyahara_JPSJ.81.023712,Miyahara_PhysRevB.89.195145} in magnetic insulators in the field of multiferroics, it can also give rise to similar physical phenomena in magnetic metals, such as the current-induced magnetization~\cite{Yanase_JPSJ.83.014703,Hayami_PhysRevB.90.024432,Hayami_doi:10.7566/JPSJ.84.064717,thole2018magnetoelectric,Watanabe_PhysRevB.98.220412,Sato_PhysRevB.103.054416,Hayami_PhysRevB.105.104428}, spin-orbital-momentum locking~\cite{Hayami_PhysRevB.104.045117}, nonreciprocal transport~\cite{Kawaguchi_PhysRevB.94.235148,Watanabe_PhysRevResearch.2.043081,Watanabe_PhysRevB.104.024416,Suzuki_PhysRevB.105.075201,Yatsushiro_PhysRevB.105.155157,hayami2022nonlinear,hayami2022nonreciprocal}, and nonlinear spin Hall/Nernst effect~\cite{Kondo_PhysRevResearch.4.013186,Hayami_PhysRevB.106.024405}. 
It has been recognized that the MTD moment can be activated not only by a vortex-like spin alignment but also by a time-reversal-odd parity mixing~\cite{yatsushiro2019atomic,watanabe2019charge} and a bond current (anapole)~\cite{Hayami_PhysRevB.101.220403,Hayami_PhysRevB.102.144441,Matsumoto_PhysRevB.104.134420,Maruyama_PhysRevX.11.011021,kanasugi2022anapole} on the basis of the quantum-mechanical expressions of the augmented multipoles in atomic~\cite{hayami2018microscopic,Hayami_PhysRevB.98.165110,kusunose2020complete,Yatsushiro_PhysRevB.104.054412} and periodic systems~\cite{Gao_PhysRevB.97.134423,Shitade_PhysRevB.98.020407,Gao_PhysRevB.98.060402}.
Furthermore, the concept of the MTD moment has been extended to higher-rank quadrupole/octupole moment~\cite{hayami2022spinconductivity, Matsumoto_PhysRevB.101.224419} and electric counterparts referred to as electric toroidal multipole~\cite{Hayami_PhysRevLett.122.147602, Nasu_PhysRevB.105.245125,hayami2021electric}, which also exhibit intriguing but different cross-correlated phenomena. 

For further understanding of the role of the MTD moment and exploring MTD-driven physical phenomena in metals, it is important to closely examine the behavior of the MTD-hosting materials. 
However, such metallic materials are still limited as compared to conventional antiferromagnetic (AFM) materials. 
The most typical example is the partially-disordered (PD) AFM metal UNi$_4$B~\cite{Mentink1994,Oyamada2007}, where the current-induced magnetization~\cite{saito2018evidence} and nonlinear Hall effect~\cite{ota2022zero} have been observed. 
Meanwhile, it is still controversial whether the MTD moment plays an essential role in these phenomena, since the details of crystal and magnetic structures have not been fully determined~\cite{Haga,tabata2021x, Willwater_PhysRevB.103.184426, Yanagisawa_PhysRevLett.126.157201}. 
Thus, it is highly desired to investigate other candidate materials to gain further insight into MTD physics.

Motivated by these situations, we focus on another AFM metal Ce$_3$TiBi$_5$ with a hexagonal structure (space group $P6_3/mcm$)~\cite{motoyama2018magnetic}, as the recent experiment observed the linear ME effect below the AFM transition temperature ($T_{\rm N}=5$~K)~\cite{shinozaki2020magnetoelectric,shinozaki2020study}. 
Although its current and magnetic field dependence implies the emergence of the in-plane MTD moment under the AFM order, the magnetic structure has not been clarified yet. 
In addition, the material exhibits a mysterious temperature dependence of magnetic susceptibility~\cite{motoyama2018magnetic}; the in-plane susceptibility ($H\perp c$) is much larger than the out-of-plane easy-axis one ($H\parallel c$), and only the out-of-plane susceptibility shows the cusp behavior at $T_{\rm N}$. 
The in-plane susceptibility keeps increasing while decreasing temperature even below $T_{\rm N}$. 

In the present study, we propose a magnetic structure that resolves the above experimental observations based on a minimal spin model. 
We find that the PD state consisting of AFM sites and nonmagnetic sites in a hexagonal cluster naturally gives rise to a nonzero ferroic MTD moment causing the linear ME effect. 
We show that magnetic frustration arising from the competition between the nearest-neighbor and further-neighbor anisotropic exchange interactions is a key ingredient to stabilize the PD state with the MTD moment within the mean-field calculations. 
Moreover, we demonstrate that the obtained PD state exhibits qualitatively similar temperature dependence to the observed magnetic susceptibility. 
We also discuss the important hopping and spin-orbit coupling (SOC) to induce the cross-correlated phenomena and quantum transport under the ferroic MTD moment. 

\begin{figure}[t!]
\begin{center}
\includegraphics[width=1.0 \hsize ]{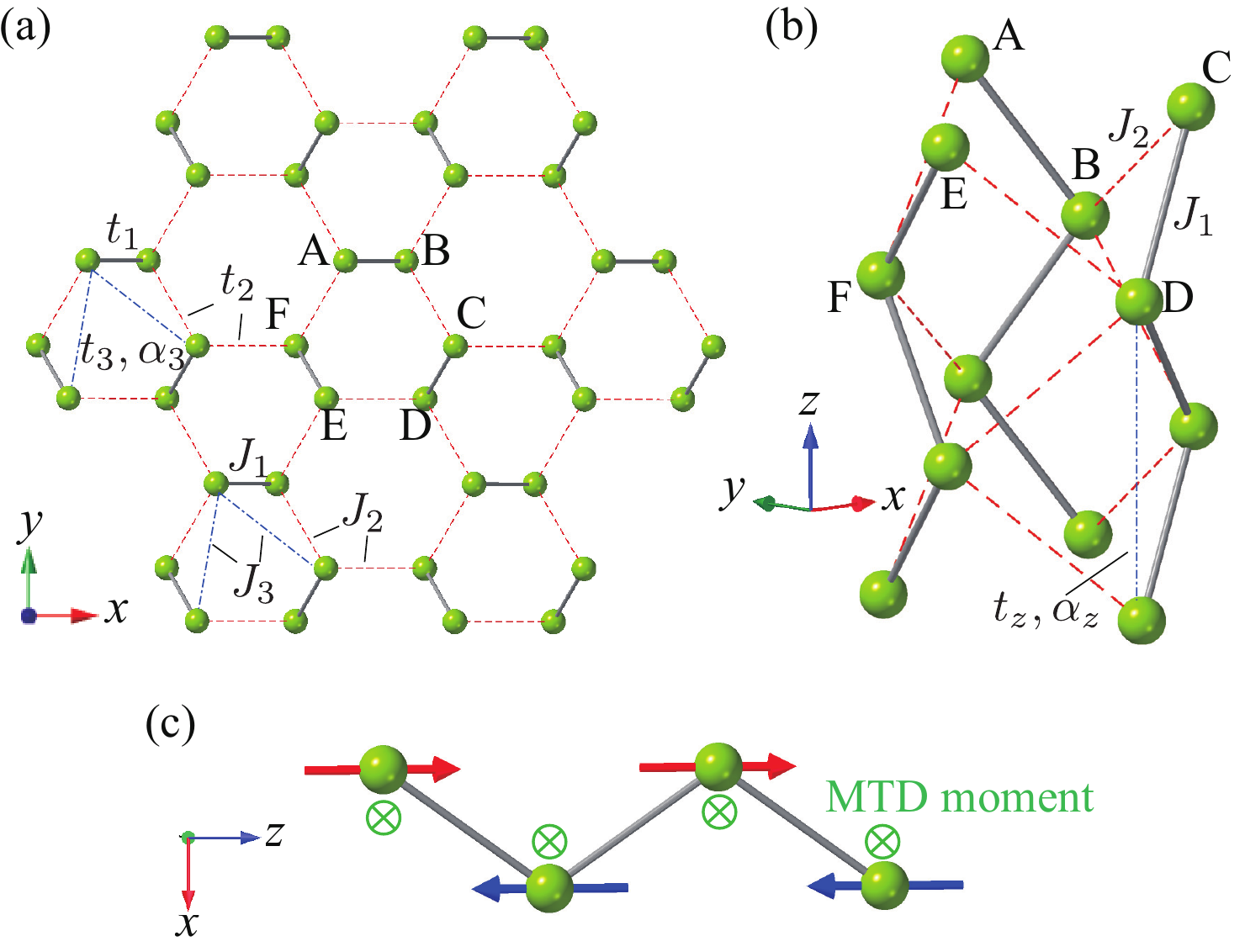} 
\caption{
\label{fig: Lattice}
(Color online) 
(a) Six-sublattice hexagonal crystal structure viewed from the $z$ direction. 
The six sublattices are labeled as A-F. 
(b) The crystal structure from a different view. 
The exchange interaction parameters, $J_1$, $J_2$, and $J_3$, in the spin model in Eq.~(\ref{eq: Ham_loc}) and the hopping parameters, $t_1$, $t_2$, $t_3$, $t_z$, $\alpha_3$, and $\alpha_z$, in the tight-binding model in Eq.~(\ref{eq: Ham_tb}) are also shown. 
(c) Staggered magnetic ordering with the $z$-spin polarization on the zigzag chain along the $z$ direction, which accompanies the uniform magnetic toroidal dipole (MTD) moment along the $y$ direction. 
}
\end{center}
\end{figure}

Let us start by considering the relationship between the AFM structure and MTD moment in Ce$_3$TiBi$_5$ under the space group $P6_3/mcm$. 
There are six Ce sites in the unit cell, which are denoted by A-F as shown in Fig.~\ref{fig: Lattice}(a). 
The sites A, C, and E (or B, D, and F) are located in the same $xy$ plane, where the sites A and B (or C and D or E and F) are related by the glide symmetry so as to form the zigzag chain along the $z$ direction, as shown in Fig.~\ref{fig: Lattice}(b). 
The three zigzag chains are related to each other by the threefold rotational symmetry.

When the staggered AFM ordering occurs in the zigzag chain with the lack of local inversion symmetry in each site, the ferroic MTD moment also emerges~\cite{Yanase_JPSJ.83.014703, Hayami_PhysRevB.90.024432, Hayami_PhysRevB.90.081115, Hayami_doi:10.7566/JPSJ.85.053705,hayami2016emergent, Hayami_PhysRevB.97.024414}. 
For example, focusing on the AB-site zigzag chain along the $z$ direction, the local crystalline electric field is present in a staggered way along the $x$ direction, i.e., $\bm{V}_{\rm A}=-\bm{V}_{\rm B} \parallel \hat{\bm{x}}$. 
Once the staggered AFM with the $z$-directional N\'eel vector, $\bm{m}_{\rm A}=-\bm{m}_{\rm B} \parallel \hat{\bm{z}}$ occurs, both spatial inversion and time-reversal symmetries are broken, which leads to the ferroic MTD moment along the $y$ direction, $T^y_{\rm A, B}\propto (\bm{V}_{\rm A, B} \times \bm{m}_{\rm A,B})^y$, as shown in Fig.~\ref{fig: Lattice}(c). 
When reversing the sign of the N\'eel vector, the sign of the MTD moment is also reversed.

\begin{figure}[t!]
\begin{center}
\includegraphics[width=1.0 \hsize ]{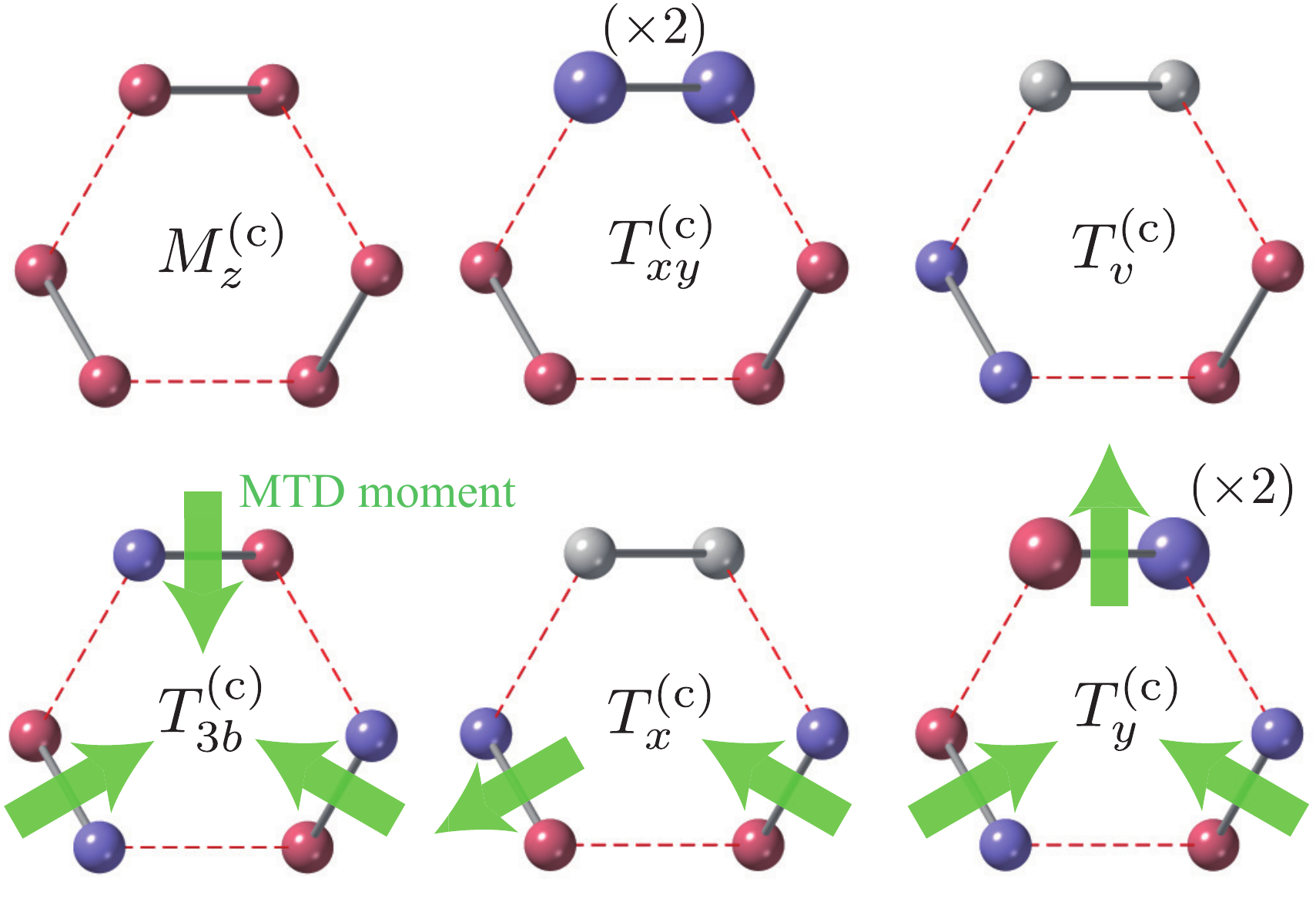} 
\caption{
\label{fig: cluster}
(Color online) 
Relation between the spin configurations and cluster multipoles in the $\bm{q}=\bm{0}$ structure.  
The color and size of the spheres represent the direction and magnitude of spins; the red, blue, and gray spheres represent the up-, down-, and zero-spin moments, respectively. 
The upper and lower three spin configurations correspond to the even- and odd-parity magnetic-type multipoles, respectively, the latter of which exhibits the MTD-moment distributions denoted by the green arrows in the three zigzag chains. 
}
\end{center}
\end{figure}

Since there are three independent zigzag chains in the unit cell, a net MTD moment in the whole system is described by their summation. 
Hereafter, we suppose that the spin moment direction lies along the $z$ direction according to the clear cusp behavior at $T_{\rm N}$ in the magnetic susceptibility~\cite{motoyama2018magnetic}. 
We also consider the $\bm{q}=\bm{0}$ magnetic structure for simplicity. 
In such a situation, six independent magnetic structures are defined, as shown in Fig.~\ref{fig: cluster}; the red, blue, and gray colors of the spheres represent the up-, down-, and zero-spin moments, respectively, and the size of the spheres represents the magnitude of spins. 
Following the manner based on cluster multipole theory~\cite{Suzuki_PhysRevB.95.094406, Suzuki_PhysRevB.99.174407}, we adapt these six magnetic structures to cluster multipoles. 
When the spin moments point in the same direction in the zigzag chain, even-parity cluster multipoles corresponding to a magnetic dipole $M^{\rm (c)}_z$ and two magnetic toroidal quadrupoles $T^{\rm (c)}_{xy}$ and $T^{\rm (c)}_{v}$ appear; no MTD moment is activated in the three zigzag chains.
On the other hand, when the spin moments point in the opposite direction so as to activate the local MTD moment in the zigzag chain, the odd-parity cluster multipoles corresponding to a magnetic toroidal octupole $T^{\rm (c)}_{3b}$ and two MTDs $T^{\rm (c)}_x$ and $T^{\rm (c)}_y$ are realized depending on the spatial distribution of the local MTD moment. 
Although there is no net MTD moment for $T^{\rm (c)}_{3b}$ owing to the cancellation of the local MTD moment in the three zigzag chains, a net in-plane MTD moment remains for $T^{\rm (c)}_x$ and $T^{\rm (c)}_y$. 
Intriguingly, such an emergence of the net MTD moment gives rise to the linear ME effect, as recently observed in experiments~\cite{shinozaki2020magnetoelectric,shinozaki2020study}. 
Thus, there are two candidate AFM structures to account for the linear ME effect in Ce$_3$TiBi$_5$: 
One is the PD state ($T^{\rm (c)}_x$) consisting of a nonmagnetic chain and two staggered AFM chains, and the other is the ferri-type state ($T^{\rm (c)}_y$ ) consisting of three staggered AFM chains with different magnitudes. 
It is noted that there is no longer threefold rotational symmetry in these states; a measurement sensitive to the lowering of the lattice symmetry like the ultrasonic measurement might be desired.

\begin{figure}[t!]
\begin{center}
\includegraphics[width=1.0 \hsize ]{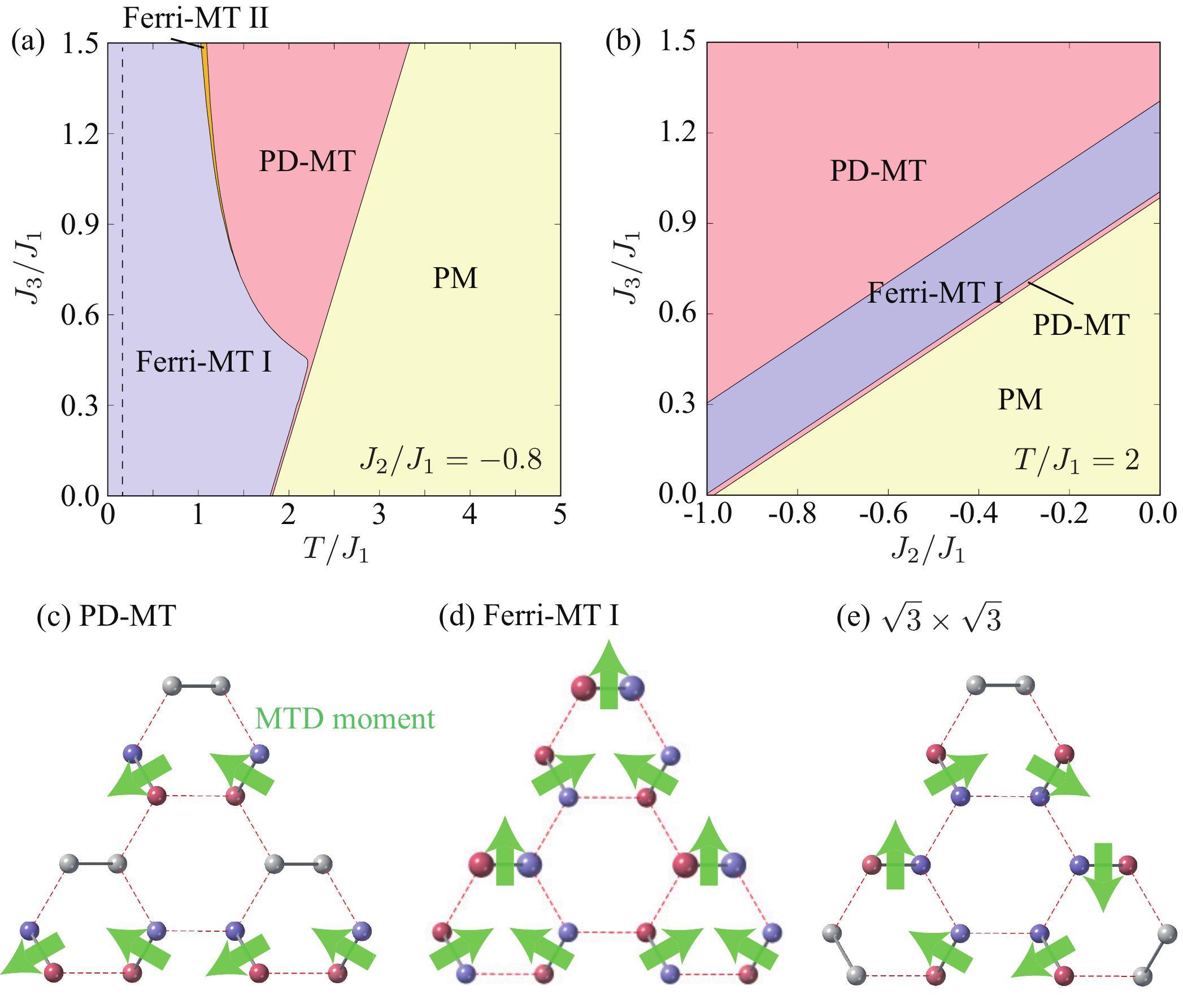} 
\caption{
\label{fig: PD}
(Color online) 
(a) $T$-$J_3$ phase diagram of the model in Eq.~(\ref{eq: Ham_loc}) for $J_2/J_1=-0.8$, which is obtained by the self-consistent mean-field calculation.
PM represents the paramagnetic state. 
The dashed line in the Ferri-MT I region represents the fully-polarized moment in each site. 
(b) $J_2$-$J_3$ phase diagram for $T/J_1=2$. 
(c), (d) The optimal spin configurations for (c) the PD-MT and (d) the Ferri-MT I; the former corresponds to $T^{\rm (c)}_x$ and the latter corresponds to a linear combination of $T^{\rm (c)}_{3b}$ and $T^{\rm (c)}_y$ in Fig.~\ref{fig: cluster}. 
The MTD-moment distribution in each state is also shown by the green arrows. 
(e) The $\sqrt{3}\times \sqrt{3}$ structure for the PD-MT state without a net MTD moment.  
}
\end{center}
\end{figure}

Next, we construct a minimal spin model to realize the AFM structures with $T^{\rm (c)}_x$ or $T^{\rm (c)}_y$. 
We consider the spin Hamiltonian with the competing exchange interactions, which is given by 
\begin{align}
\label{eq: Ham_loc}
\mathcal{H}^{\rm loc}= &J_1\sum_{\langle ij \rangle} S_i^z S_j^z
+J_2\sum_{\langle \langle ij \rangle \rangle} S_i^z S_j^z
+J_3\sum_{\langle\langle\langle ij \rangle\rangle\rangle} S_i^z S_j^z,
\end{align}
where $\bm{S}_i=(S_i^x, S_i^y, S_i^z)$ is the classical spin at site $i$ with $|\bm{S}_i|=1$. 
$J_1$, $J_2$, and $J_3$ stand for the Ising-type coupling constants between the nearest-, second-nearest-, and third-nearest-neighbor spins, respectively; see also Figs.~\ref{fig: Lattice}(a) and \ref{fig: Lattice}(b).  
We set $J_1$ to be AFM ($J_1>0$) to induce the local MTD moment in each zigzag chain. 
Owing to the bi-partite network, the magnetic structure with $T^{\rm (c)}_{3b}$ is stabilized for the AFM $J_2$ and ferromagnetic (FM) $J_3$. 
Thus, we consider the FM $J_2<0$ and the AFM $J_3$ in order to enhance the instability toward the magnetic structures with $T^{\rm (c)}_x$ and $T^{\rm (c)}_y$. 
In other words, magnetic frustration is important to stabilize these structures, which has been also implied in experiments~\cite{tsubouchi2020temperature}. 
In the following, we take $J_1$ to be the energy unit of the model.

\begin{figure}[t!]
\begin{center}
\includegraphics[width=1.0 \hsize ]{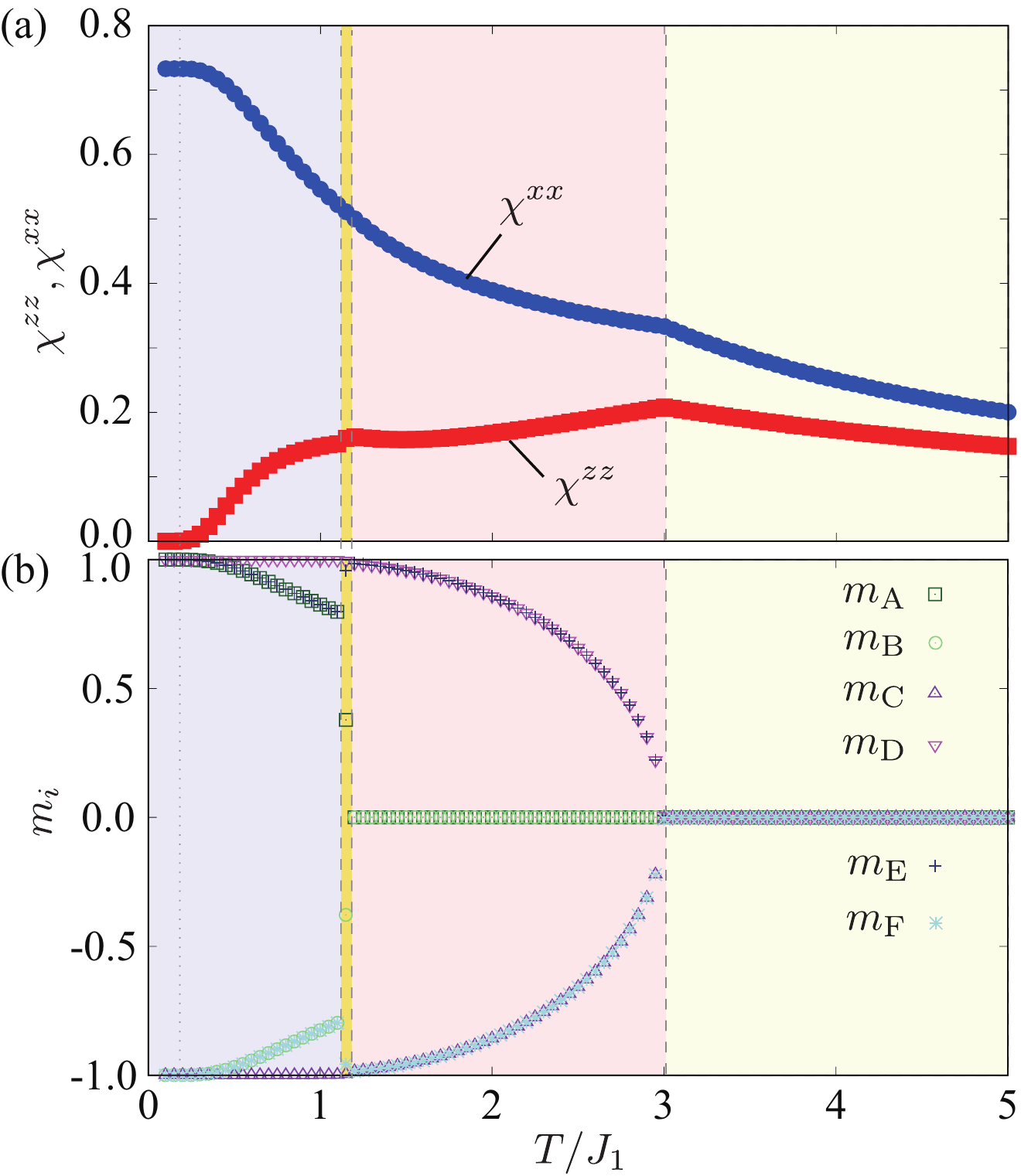} 
\caption{
\label{fig: suscep}
(Color online) 
Temperature ($T$) dependence of (a) magnetic susceptibility, $\chi^{xx}$ and $\chi^{zz}$, and (b) averaged spin moments at site $i=$A-F, $m_i$, at $J_2/J_1=-0.8$ and $J_3/J_1=1.2$. 
The dashed and dotted lines stand for the phase boundaries and the fully-polarized boundary, respectively. 
}
\end{center}
\end{figure}

Figure~\ref{fig: PD}(a) shows the phase diagram while changing the temperature $T$ and $J_3$ for fixed $J_2/J_1=-0.8$. 
The phase diagram is constructed by performing the self-consistent mean-field calculations for the six-sublattice system under the periodic boundary conditions. 
In the intermediate $T$ region for $J_3/J_1 \gtrsim 0.44$, we find that the AFM structure with $T^{\rm (c)}_x$ is robustly stabilized in Fig.~\ref{fig: PD}(a), where we refer to this phase as PD-MT.  
The schematic spin configuration is shown in Fig.~\ref{fig: PD}(c).  
In this phase, four out of the six sites have nonzero expectation values of the $z$-polarized spin moment $m_i=\braket{S^z_{i}}$; the moments are developed while decreasing $T$ from the paramagnetic state, as shown in the case of $J_3/J_1=1.2$ in Fig.~\ref{fig: suscep}(b). 
The appearance of the PD state in the intermediate $T$ region is common to other frustrated spin models~\cite{Mekata_JPSJ.42.76,wada1982monte,fujiki1983monte,Landau_PhysRevB.27.5604,takayama1983monte,fujiki1984possibility,takagi1995new,todoroki2004ordered}.

While decreasing $T$ from the PD-MT, two ferrimagnetic phases with the in-plane MTD moment (Ferri-MT I and II) appear by having nonzero $m_i$ in the remaining nonmagnetic site, as shown in Fig.~\ref{fig: PD}(a). 
The Ferri-MT I is characterized by two types of inequivalent zigzag chains to possess the MTD moment along the $y$ direction, whose structure is shown in Fig.~\ref{fig: PD}(d). 
This spin configuration is characterized by a linear combination of $T^{\rm (c)}_{3b}$ and $T^{\rm (c)}_{y}$ in Fig.~\ref{fig: cluster}. 
The other Ferri-MT II appears in the region sandwiched by the PD-MT and Ferri-MT I, where the spin magnitudes in the three zigzag chains are different from each other, as shown in Fig.~\ref{fig: suscep}(b). 
For small $J_3/J_1 \lesssim 0.44$, almost the direct phase transition from the paramagnetic state to the Ferro-MT I occurs~\cite{comment_ferri_PD}.

We show the $J_2$ and $J_3$ dependence of the stability of the PD-MT and Ferri-MT I for fixed $T/J_1=2$ in Fig.~\ref{fig: PD}(b). 
The result indicates that larger FM $J_2$ and AFM $J_3$, which tend to enhance magnetic frustration, are favored to realize both states.  
In particular, larger frustration is required to realize the PD-MT.

Let us comment on the degeneracy of the obtained spin configurations. 
As the sites A, C, and E (or B, D, and F) in the same $xy$ plane form the distorted kagome network, the present three-sublattice PD-MT is degenerate with the other spin configurations characterized by a finite-$q$ long-period modulation, such as the $\sqrt{3}\times \sqrt{3}$ structure [see Fig.~\ref{fig: PD}(e)] within the interactions $J_1$-$J_3$; the latter $\sqrt{3}\times \sqrt{3}$ structure does not have a net MTD moment owing to the cancellation. 
Thus, one needs additional further-neighbor interactions and/or magnetic anisotropy to lift their degeneracy, but according to the experimental observation of the linear ME effect~\cite{shinozaki2020magnetoelectric,shinozaki2020study}, we here implicitly assume that the $\bm{q}=\bm{0}$ structure (PD-MT) is stabilized by such an effect.

Next, we discuss the behavior of magnetic susceptibility in the PD-MT state. 
Figure~\ref{fig: suscep}(a) shows the $T$ dependence of longitudinal and transverse susceptibility, $\chi^{zz}$ and $\chi^{xx}$, at $J_2/J_1=-0.8$ and $J_3/J_1=1.2$. 
We find two characteristic features in their $T$ dependence: One is the increase of $\chi^{xx}$ while decreasing $T$ even below $T_{\rm N}$ ($T_{\rm N}/J_1 = 3$) like the paramagnetic state and the other is the relation $\chi^{xx}>\chi^{zz}$ in the whole $T$ region in spite of the $z$-polarized AFM ordering. 
The former is attributed to inequivalence among three zigzag chains. 
The latter is explained by the difference between the effective FM interactions of three zigzag chains in the system; the condition for $\chi_{xx} >\chi_{zz}$ is given by $J_1+2J_2+2J_3>0$ in the present model. 
As these two features are consistent with the experimental observations in Ce$_3$TiBi$_5$~\cite{motoyama2018magnetic}, it is expected that the interplay between the weak FM $xy$-spin interaction and the strong AFM $z$-spin one is a key factor in this coumpound~\cite{comment_uud}.

As discussed above, the PD-MT state appears only for finite $T$, where the ground-state spin configuration is replaced by the Ferri-MT I. 
In addition, the PD-MT state in the classical spin model might be a quasi-long-range ordering by performing unbiased Monte Carlo simulations~\cite{wada1982monte,fujiki1983monte,Landau_PhysRevB.27.5604,takayama1983monte,fujiki1984possibility,takagi1995new,todoroki2004ordered}. 
Nevertheless, it is noteworthy to mention that it can remain as a stable state by considering the spin-charge coupling~\cite{Ishizuka_PhysRevLett.108.257205} and local Kondo singlet formation~\cite{LacroixUNi4B,Motome2010,Hayami2011,Hayami2012,Hayami_Conference,Aulbach_PhysRevB.92.235131,Hayami_1742-6596-592-1-012101,Kessler_PhysRevB.102.235125} in the presence of itinerant electrons. 
Especially, the latter case ensures the stabilization of the PD-MT even at zero temperature, as indeed found in a different Ce-based compound, CePdAl~\cite{Donni, Oyamada2008, Zhao_PhysRevB.94.235131, Lucas_PhysRevLett.118.107204,mochidzuki2017thermodynamic}.  
To resolve whether another phase transition occurs for a lower $T$ region, further magnetic susceptibility measurements are required.

So far, we have investigated the AFM structures within the localized spin model. 
As the obtained structures have nonzero MTD moments, their related physical phenomena are expected. 
To this end, we consider the important hopping and SOC parameters to induce such physical phenomena by additionally considering the tight-binding Hamiltonian given by 
\begin{align}
\label{eq: Ham_tb}
\mathcal{H}^{\rm tb}= &\sum_{ij \sigma} t_{ij} c^{\dagger}_{i\sigma} c^{}_{j\sigma} + \sum_{ij \sigma\sigma'} \bm{g}_{ij}\cdot c^{\dagger}_{i\sigma} \bm{\sigma}_{\sigma\sigma'} c^{}_{j\sigma'},  \end{align}
where $c^{\dagger}_{i\sigma}$ ($c^{}_{i\sigma}$) is the creation (annihilation) operator of electrons with site $i$ and spin $\sigma$. 
The first term represents the hoppings between sites $i$ and $j$. 
We adopt four hopping parameters: $t_1$, $t_2$, and $t_3$ between the different sublattices [Fig.~\ref{fig: Lattice}(b)] and $t_z$ between the same sublattice [Fig.~\ref{fig: Lattice}(c)]. 
The second term represents the spin-dependent hoppings that originate from the relativistic SOC in the absence of local inversion symmetry of the zigzag chain; we consider the out-of-plane hopping with the in-plane spin component between the same sublattice, $\bm{g}_{ij}= 2 \sin k_z (g^x_{ii},g^y_{ii},0)=2\sin k_z \bm{g}^{\parallel}_\eta$ ($\eta=$ A-F) and $|\bm{g}^{\parallel}_\eta|=\alpha_z$, and the in-plane hopping with the out-of-plane spin component between the different sublattices, $\bm{g}_{ij}=(0,0,g^{\perp}_{ij})$ and $|g^{\perp}_{ij}|=\alpha_3$, satisfying the sixfold rotational symmetry of the lattice structurer.

For the total Hamiltonian $\mathcal{H}^{\rm loc}+\mathcal{H}^{\rm tb}$ in Eqs.~(\ref{eq: Ham_loc}) and (\ref{eq: Ham_tb}), we calculate the essential hopping and SOC to cause the MTD-related physical phenomena under the PD-MT: asymmetric band-bottom shift $\varepsilon^{\rm A}(\bm{k})$, linear ME effect $\alpha_{\mu;\nu}$ ($M_{\nu}=\alpha_{\mu;\nu} E_{\mu}$), Drude-type nonlinear current conductivity $\sigma_{\mu;\nu\eta}$ ($J_{\mu}=\sigma_{\mu;\nu\eta} E_{\nu}E_{\eta}$), and nonlinear spin current conductivity $\sigma^{\rm (s)}_{\mu;\nu\eta}$ ($J^{\rm (s)}_{\mu}=\sigma^{\rm (s)}_{\mu;\nu\eta} E_{\nu}E_{\eta}$)~\cite{Hayami_PhysRevB.106.024405} for $\mu,\nu,\eta=x,y,z$; $\bm{E}$, $\bm{M}$, $\bm{J}$, and $\bm{J}^{\rm (s)} \equiv \bm{J}S^z$ represent the electric field, magnetization, electric current, and $z$-polarized spin current, respectively. 
Their parameter dependence is obtained by expanding the Hamiltonian and the response tensors as follows~\cite{Hayami_PhysRevB.102.144441,Oiwa_doi:10.7566/JPSJ.91.014701}: $\varepsilon^{\rm A}(\bm{k}) \propto {\rm Tr}[H^i(\bm{k})]-{\rm Tr}[H^i(-\bm{k})]$, $\alpha_{\mu\nu} \propto \sum_{\bm{k}}{\rm Tr}[\sigma^\mu H^i(\bm{k})v_{\nu\bm{k}}H^j (\bm{k})]$, $\sigma_{\mu;\nu\eta} \propto \sum_{\bm{k}}{\rm Tr}[v_{\mu\bm{k}}H^i(\bm{k})v_{\nu\bm{k}}H^j (\bm{k})v_{\eta\bm{k}}H^k (\bm{k})]$, and $\sigma^{\rm (s)}_{\mu;\nu\eta} \propto \sum_{\bm{k}}{\rm Tr}[v_{\mu\bm{k}}\sigma^z H^i(\bm{k})v_{\nu\bm{k}}H^j (\bm{k})v_{\eta\bm{k}}H^k (\bm{k})]$, where $H^i (\bm{k})$ is the $i$th power of the Hamiltonian matrix at wave vector $\bm{k}$, $\bm{v}_{\bm{k}}=\partial H(\bm{k})/\partial \bm{k}$ is the current operator, and $\bm{\sigma}$ is the spin operator.

\begin{table}[h]
  \caption{
  Essential spin-orbit coupling (SOC) for the asymmetric band-bottom shift and physical tensors in the PD-MT phase.
     }
  \label{tab}
  \begin{center}
    \begin{tabular}{lllcccccc}
      \hline \hline
      &                                                               & SOC   \\ \hline
   band-bottom shift &  $\varepsilon^{\rm A}(\bm{k})$   &  $ \alpha_{3} $         \\       \hline
  linear ME  &   $\alpha_{y;z}$   &  $\alpha_z $    \\ 
   &   $\alpha_{z;y}$   &  $\alpha_z^2 $    \\ \hline
  nonlinear current  &   $\sigma_{x;xx}$   &  $ \alpha_{3} $     \\
   &   $\sigma_{x;yy}=\sigma_{y;xy}$   &  $\alpha_3 $    \\ 
      &   $\sigma_{x;zz}=\sigma_{z;xz}$   & $\alpha_3$ or $\alpha_z$     \\ \hline
  nonlinear spin current  &   $2\sigma^{\rm (s)}_{y;xx}=-\sigma^{\rm (s)}_{x;xy}$   &  No     \\
      &   $2\sigma^{\rm (s)}_{y;zz}=-\sigma^{\rm (s)}_{z;yz}$   &  No     \\
      \hline \hline
    \end{tabular}
  \end{center}
\end{table}

The results are summarized in Table~\ref{tab}, which indicates that the different SOC dependence is found in different response tensors. 
The in-plane SOC $\alpha_3$ is necessary to induce the asymmetric band deformation   $\varepsilon^{\rm A}(\bm{k})$ and the nonlinear nonreciprocal current $\sigma_{x;xx}$ and $\sigma_{x;yy}=\sigma_{y;xy}$, while the out-of-plane one $\alpha_{z}$ plays a role in causing the linear ME effect $\alpha_{y;z}$ and $\alpha_{z;y}$. 
Either $\alpha_3$ or $\alpha_z$ is necessary for the transverse nonlinear current $\sigma_{x;zz}=\sigma_{z;xz}$ and no SOC is needed for the nonlinear spin current $2\sigma^{\rm (s)}_{y;xx}=-\sigma^{\rm (s)}_{x;xy}$ and $2\sigma^{\rm (s)}_{y;zz}=-\sigma^{\rm (s)}_{z;yz}$. 
As the linear ME was observed in experiments, the transverse nonlinear current $\sigma_{x;zz}=\sigma_{z;xz}$ is also expected, since both are proportional to the same SOC parameter $\alpha_z$. 
Thus, the experimental measurement of $\sigma_{x;zz}=\sigma_{z;xz}$ remains an interesting issue. 

To summarize, we have investigated the AFM structure to exhibit the linear ME effect keeping with Ce$_3$TiBi$_5$ in mind. 
Based on the experimental findings~\cite{motoyama2018magnetic,shinozaki2020magnetoelectric,shinozaki2020study}, we propose that the PD state with the MTD moment is a candidate to account for the emergence of the ME effect and the temperature dependence of the magnetic susceptibility in Ce$_3$TiBi$_5$. 
We show the important exchange interaction parameters by analyzing the spin model. 
Furthermore, we present the essential SOC to cause the physical phenomena induced by the MTD moment.

\begin{acknowledgments}
The authors thank G. Motoyama, M. Shinozaki, and T. Mutou for fruitful discussions. 
This research was supported by JSPS KAKENHI Grants Numbers JP21H01031, JP21H01037, JP22H04468, JP22H00101, JP22H01183, and by JST PRESTO (JPMJPR20L8).
Parts of the numerical calculations were performed in the supercomputing systems in ISSP, the University of Tokyo.
\end{acknowledgments}

\bibliographystyle{JPSJ}
\bibliography{ref}

\begin{thebibliography}{10}
\bibitem{comment_ferri_PD}
The appearance of the PD state in the narrow region for $J_3/J_1 \lesssim 0.44$
  might be artifact owing to the numerical accuracy.
\bibitem{comment_uud}
A similar behavior also happens for the region where the paramagnetic state
  directly changes into the Ferri-MT I. To distinguish the PD-MT and Ferri-MT
  I, the NQR/NMR and $\mu$SR measurements are preferred.
\end{thebibliography}

\end{document}